\documentclass[%
 reprint,
nofootinbib,
 amsmath,amssymb,
 aps,
]{revtex4-1}

\usepackage{graphicx}
\usepackage{dcolumn}
\usepackage{bm}
\usepackage{hyperref}

\begin{document}

\title{Stationary Black Holes with Time-Dependent Scalar Fields}
\author{Alexander A. H. Graham}
\email{A.A.H.Graham@damtp.cam.ac.uk}
 \author{Rahul Jha}%
 \email{ R.Jha@damtp.cam.ac.uk}
\affiliation{%
Department of Applied Mathematics and Theoretical Physics\\
Centre for Mathematical Sciences\\
University of Cambridge\\
Wilberforce Road, Cambridge, CB3 0WA, UK 
}%

\date{\today}

\begin{abstract}
It has been well known since the 1970s that stationary black holes do not generically support scalar hair. Most of the no-hair theorems which support this depend crucially upon the assumption that the scalar field has no time dependence. Here we fill in this omission by ruling out the existence of stationary black hole solutions even when the scalar field may have time dependence. Our proof is fairly general, and in particular applies to non-canonical scalar fields and certain non-asymptotically flat spacetimes. It also does not rely upon the spacetime being a black hole.
\begin{description}
\item[PACS numbers] 04.70.Bw, 04.50.Kd
\end{description}
\end{abstract}

\maketitle


\section{\label{sec:level1}Introduction}
It has become something of a clich\'e in the relativity community that, in the colourful language of Wheeler, black holes have no hair, meaning more precisely that stationary, asymptotically flat black hole solutions cannot generically support long-ranged scalar fields \cite{wald84, hawking73, chandrasekhar, chrusciel12, bekenstein98}. Like many clich\'es there is quite a fair amount of evidence to support this belief. The strongest evidence comes from the various no-hair results on the subject originally proven by Chase and Bekenstein in the early 1970s \cite{chase70, bekenstein72a, bekenstein72b, bekenstein72c}, and greatly extended by Sudarsky, Bekenstein, Heusler and others in the 1990s \cite{sudarsky95, bekenstein95, heusler96}. The best known results rule out the existence of stationary black holes with scalar hair for a real scalar field with self-interaction potential \(V(\phi)\) obeying \(V_{,\phi\phi}>0\) or \(\phi{}V_{,\phi}\geq0\), and static, spherically symmetric black holes with scalar hair assuming only that the potential is bounded from below. These proofs can even be extended to non-canonical scalar fields \cite{bekenstein95, graham14} and to Galileons to some extent \cite{hui13, sotiriou13}.

Despite the importance of these results, almost all the results above have a subtle assumption: they implicitly assume the scalar field has no time dependence. While seemingly reasonable this is actually not entirely justified. If the spacetime is stationary then clearly the energy-momentum tensor of the scalar field does not depend on the timelike coordinate, but this does not always imply the scalar field may have no time dependence. While for a real, canonical scalar field this is only possible with a massless scalar field \cite{hoenselaers78}, additional possibilities exist if the scalar field has a non-canonical kinetic structure. The purpose of this paper is to rule out this scenario. 

Our proof will proceed via a study of the Einstein equations and the spacetime asymptotics. In fact, we shall establish the following somewhat stronger result: no stationary, asymptotically flat, four-dimensional solutions of the Einstein equations coupled to a non-canonical, real scalar field exist if the scalar field depends upon time, modulo certain assumptions on the scalar field action which will be described later. That the solutions are black holes is not essential to the argument. It is important to note that the proof depends crucially upon the scalar field being real, and there being only one scalar field: if either assumption is relaxed the argument does not hold. Indeed, static solutions of the Einstein equations coupled to a complex scalar field have been known about for a long time in the context of boson stars \cite{liebling12}. Recently, it was claimed that a family of hairy black hole solutions coupled to a complex Klein-Gordon field were found numerically \cite{herdeiro14}. 

In this paper units are employed in which \(c=G=1\).

\section{Preliminaries}
In this paper we study four-dimensional spacetimes which are stationary. We will also make the assumption that the spacetime contains a scalar field which obeys the null energy condition. This means the rigidity theorem holds, and so the spacetime must also be axisymmetric \cite{hawking72, hollands07}. One can then show that we may choose coordinates \((t,r,\theta,\phi)\) so that the spacetime metric takes the form \cite{chandrasekhar}
\begin{align} \label{2.1}
&ds^{2}=-e^{\mu(r,\theta)}dt^{2}+2\rho(r,\theta)dtd\phi+e^{\nu(r,\theta)}d\phi^{2} \nonumber
\\
&+e^{A(r,\theta)}dr^{2}+e^{B(r,\theta)}d\theta^{2}.
\end{align}
Note that the spacetime has topology \(\mathbb{R}^{2}\times{}S^{2}\), with \(\theta\) and \(\phi\) periodic coordinates on the two-sphere. The metric is static if \(\rho=0\), and it can be easily verified by direct calculation that the following components of the Ricci tensor are zero:
\begin{equation} \label{2.2}
R_{tr}=R_{t\theta}=R_{r\phi}=R_{\theta\phi}=0. 
\end{equation}
This will be key to the argument and will be used in section III. We shall also assume that the spacetime is asymptotically flat. This implies that the metric components must decay appropriately at infinity: \(\rho{}\rightarrow0\) and \(\mu{}\rightarrow0\) as \(r\rightarrow\infty\). It also implies that the components of the energy-momentum tensor of the scalar field must tend to zero in this coordinate system as \(r\rightarrow\infty\).   

\section{Real scalar field}
We now give the argument for a real scalar field, \(\Phi\). We shall assume that the scalar field action does not depend explicitly on more than one derivative of the scalar field, but we will allow the field to have a non-canonical kinetic structure. Our proof thereby encompasses K-essence type scalar fields \cite{mukhanov99, chiba00, ap00, ap01}. We take the scalar field action to be
\begin{equation} \label{3.1}
S=\int{}d^{4}x\sqrt{-g}P(\Phi,X),
\end{equation}  
where \(X=-\frac{1}{2}\nabla_{a}\Phi\nabla^{a}\Phi\). A canonical scalar field is given by the choice \(P=X-V(\Phi)\). By varying the action with respect to the scalar field we obtain the equation of motion for the scalar field:
\begin{equation} \label{3.2}
\nabla_{a}(P_{,X}\nabla^{a}\Phi)+P_{,\Phi}=0, 
\end{equation}
where \(P_{,X}=\partial{}P/\partial{}X\) and \(P_{,\Phi}=\partial{}P/\partial{}\Phi\). We will assume that \(P_{,X}\neq0\), as otherwise the theory is classically not well posed as an initial value problem (for more details on the initial value formulation of non-canonical scalar fields see Refs. \cite{lim05, rendall06}). Similarly, varying the action with respect to the metric shows the energy-momentum tensor of the scalar field to be
\begin{equation} \label{3.3}
T_{ab}=P_{,X}\partial_{a}\Phi\partial_{b}\Phi+Pg_{ab}. 
\end{equation}

Now let us assume there does exist a stationary, asymptotically flat black hole with a non-trivial, time-dependent scalar field. We argue by contradiction that no such solutions exist. The first step of the argument makes explicit use of the Einstein equations: \(R_{ab}=8\pi{}(T_{ab}-\frac{1}{2}Tg_{ab})\). Specifically, the \(tr\) and \(t\theta\) components of the Einstein equations imply that
\begin{equation} \label{3.4}
\partial_{t}\Phi\partial_{r}\Phi=0, \ {}\ \partial_{t}\Phi\partial_{\theta}\Phi=0.
\end{equation}
Clearly then, if \(\Phi\) does depend upon time then it cannot depend upon the coordinates \(r\) and \(\theta\), so that \(\Phi=\Phi(t,\phi)\). This is the key step in the argument. We note that the argument would apply equally well if the cosmological constant were non-zero, or indeed if \(P_{,\Phi}\neq0\). It is a generalisation of an old observation of Wyman \cite{wyman81}, who studied static, spherically symmetric solutions of the Einstein equations coupled to a massless scalar field \(\phi\): the Einstein equations in this case take the form \(R_{ab}=8\pi\partial_{a}\phi\partial_{b}\phi\). While no black hole solutions exist if the scalar field is non-constant the general solution when the scalar field is purely radial has been known for a long time, having been discovered first by Fisher \cite{fisher48} and rediscovered several times since then (see the translators prefix to the arXiv version of Ref. \cite{fisher48}). It describes in general a naked singularity and is normally known as the Janis-Newman-Winicour solution \cite{janis68}. Wyman realised that in principle the scalar field could also depend upon time, but that the \(tr\) component of the Einstein equations implied either that \(\phi=\phi(r)\) or \(\phi=\phi(t)\).

The next step of the argument comes from inspection of the energy-momentum tensor, which quickly reveals that this situation can only occur in general if the Lagrangian has no explicit dependence on \(\Phi\), meaning that \(P=P(X)\). Furthermore, the scalar field can only depend linearly upon time. This is because if either of these assumptions were false then some components of the energy-momentum tensor will depend explicitly upon time, which would preclude the spacetime geometry from being stationary. More precisely, the \(T_{tt}\) and \(T_{rr}\) components of the energy-momentum tensor imply that \(P\) and \(P_{,X}\dot{\Phi}^{2}\) are independent of time. In the canonical case this immediately implies that \(\Phi\) is linear in \(t\), and that there is no non-constant potential term. In the non-canonical case one solution is given by \(\Phi\) being linear in \(t\), which implies \(P=P(X)\). For general actions this is the only solution, as there are two equations to be satisfied for \(\Phi\) so the system is overdetermined\footnote{There may be some choices where this is not true, but they will in general be artificial. On possible example would be actions of the form \(P=\alpha\ln{}X+P_{0}(\phi)\), where \(\alpha\) is a constant, but such actions are in general unacceptable. In particular, this is not compatible with asymptotic flatness.}.

Actually, we can also rule out \(\Phi\) depending upon \(\phi\). This is because the spacetime is axisymmetric, so a similar argument to the above implies that \(T_{ab}\) can only be consistent with axisymmetry of the spacetime geometry if \(\Phi\) depends at most linearly upon \(\phi\). However, since \(\phi\) is a periodic coordinate clearly this cannot be so, as otherwise \(\Phi\) would not be a continuous, single-valued function. Hence the scalar field can in fact only depend upon time.

We have now deduced that the only scalar field consistent with this geometry is
\begin{equation} \label{3.45}
\Phi=\alpha{}t+\beta,  
\end{equation}
where \(\alpha\) and \(\beta\) are constants. Clearly this is a highly degenerate case, which we would hardly expect to be compatible with the above geometry. The final step in the argument is to show this is correct. More precisely, we will show that this is in general only consistent with asymptotic flatness if \(\alpha=0\). This can be seen by considering the asymptotic value of the energy-momentum tensor at infinity. Since \(g^{tt}\rightarrow-1\) as \(r\rightarrow\infty\) then \(X\rightarrow\alpha^{2}/2\), and the \(tt\) and \(rr\) components of the energy-momentum tensor tend to
\begin{equation} \label{3.5}
T_{tt}\rightarrow{}P_{,X}(\alpha^{2}/2)\alpha^{2}-P(\alpha^{2}/2),\ {}\ T_{rr}\rightarrow{}P(\alpha^{2}/2). 
\end{equation}
If the spacetime is asymptotically flat these must tend to zero (the vanishing of these components of the energy-momentum tensor suffices to establish the vanishing of all other components of the energy-momentum tensor at infinity), but generically they tend to a non-zero constant if \(\alpha\neq0\). Indeed, for them both to tend to zero with \(\alpha\neq0\) the action must satisfy 
\begin{equation} \label{3.6}
P(\alpha^{2}/2)=0,\ {}\ P_{,X}(\alpha^{2}/2)=0. 
\end{equation}
Since these compose two equations for one unknown the system is overdetermined, and in general no solutions exist. In particular, these conditions are not satisfied in the canonical case so scalar hair for a massless, canonical scalar field is entirely ruled out. It should also be noted that a theory in which \( P_{,X}(\alpha^{2}/2)=0\) at infinity is likely to be pathological to some degree, since one normally requires \(P_{,X}>0\) for the theory to be well defined. Thereby we conclude that, neglecting some highly pathological cases which satisfy Eq. \eqref{3.6} when \(\alpha\neq0\), the black hole cannot support time-dependent scalar hair.

\section{(anti-) de Sitter boundary conditions}
Although our main focus is on solutions which are asymptotically flat let us briefly consider the case when the spacetime is either asymptotically anti-de Sitter or de Sitter \cite{torii99, torii01}. This will occur either if the cosmological constant is non-zero or the scalar field has a constant term in the action (needless to say there is a degeneracy between these two concepts). The first thing to note is that this modifies only the last stage of the argument. In particular, the Einstein equations still imply that \(\Phi=\Phi(t,\phi)\) even when \(\Lambda\neq0\), and we must have that \(\Phi=\alpha{}t+\beta\), where \(\alpha\) and \(\beta\) are constants.

The case of an asymptotically anti-de Sitter spacetime is somewhat more straightforward. In this case then asymptotically \(g^{tt}\rightarrow0\) from below as \(r\rightarrow\infty\). This implies that as \(r\rightarrow\infty\) \(X\rightarrow0\) and the \(tt\) and \(rr\) components of the energy-momentum tensor tend to 
\begin{equation} \label{4.5}
T_{tt}\rightarrow{}P_{,X}(0)\alpha^{2}-\left(1+\frac{|\Lambda|r^{2}}{3}\right)P(0),\ {}\ T_{rr}\rightarrow{}0. 
\end{equation}
For the spacetime to be asymptotically anti-de Sitter we require that these components of the energy-momentum tensor tend to zero as \(r\rightarrow\infty\). This requires that \(P(0)=0\), and either \(\alpha=0\) or that \(P_{,X}(0)=0\). Clearly, the first and third conditions cannot be satisfied generically, and so we will normally require that \(\alpha=0\) and there is no time-dependent scalar hair. This in particular applies to a canonical scalar field.

The case of an asymptotically de Sitter spacetime is somewhat different. These spacetimes possess a cosmological event horizon at \(r=r_{c}\). Now since \(g^{tt}\) diverges at the cosmological horizon in this coordinate system it is clear that the scalar \(X\rightarrow\infty\) as \(r\rightarrow{}r_{c}\) unless \(\alpha=0\). However, it is also clear that in general this is incompatible with the spacetime being asymptotically de Sitter, for in general we expect that \(P\rightarrow\pm\infty\) as \(X\rightarrow\infty\) and so all the components of the energy-momentum tensor will diverge at the cosmological horizon. Note this is not an artefact of the coordinate system failing to cover the horizon; even if we used Eddington-Finkelstein like coordinates to cover the horizon all components of \(T_{ab}\) not identically zero would diverge as \(r\rightarrow{}r_{c}\) due to the divergence of \(P\). Hence we must in general have \(\alpha=0\), and so the black hole possesses no scalar hair. Again, this in particular applies to a canonical scalar field.

\section{Complex scalar field}
It is natural to wonder if the argument given above also works for a complex scalar field, \(\Psi\). After all, the original no-hair theorem of Bekenstein generalises fairly straightforwardly to a complex scalar field (this is also the case for the results involving a non-canonical scalar field presented in Ref. \cite{graham14}). We will therefore consider this case briefly. As before, we will restrict the scalar field action to be first order in the derivatives of the field but allow it to be in principle non-canonical. Since the action must be real we take it to be of the form
\begin{equation} \label{5.1}
S=\int{}d^{4}x\sqrt{-g}P(|\Psi|^{2},Y),
\end{equation}  
where \(Y=-\nabla_{a}\Psi^{*}\nabla^{a}\Psi\) and \(|\Psi|^{2}=\Psi^{*}\Psi\). A canonical complex scalar field is given by the action \(P=Y-V(|\Psi|^{2})\). The equations of motion are easily found to be
\begin{equation} \label{5.2}
\nabla_{a}(P_{,Y}\nabla^{a}\Psi)+P_{,|\Psi|^{2}}\Psi=0, 
\end{equation}
and the energy-momentum tensor is given by
\begin{equation} \label{5.3}
T_{ab}=2P_{,Y}\partial_{(a}\Psi^{*}\partial_{b)}\Psi+Pg_{ab}. 
\end{equation}
It is now fairly easy to see why the argument of section III does not apply in this case. Firstly, it is no longer the case that we require \(P_{,|\Psi|^{2}}=0\) if the energy-momentum tensor is to be independent of time; even for a canonical scalar field a potential of the form \(V(|\Psi|^{2})\) can be compatible with \(\Psi\) having some time dependence and the spacetime being stationary. More importantly, use of the Einstein equations does not imply that \(\Psi\) may not depend upon \(r\) or \(\theta\). In fact, the \(tr\) and \(t\theta\) components of the Einstein equations show that
\begin{equation} \label{5.4}
\partial_{(t}\Psi^{*}\partial_{r)}\Psi=0, \ {}\ \partial_{(t}\Psi^{*}\partial_{\theta)}\Psi=0,
\end{equation}
which no longer implies \(\partial_{r}\Psi=0\) and \(\partial_{\theta}\Psi=0\) if \(\partial_{t}\Psi\neq0\).

\section{Conclusion}
In this paper we investigated whether it is possible that there exists any stationary, asymptotically flat black holes possessing time-dependent scalar hair and showed that this is generically impossible, and is completely excluded for a canonical scalar field. This completes an important omission in the original no-hair theorems. Our proof followed fairly straightforwardly via an analysis of the off-diagonal components of the Einstein equations. Since our proof only uses a small subset of the field equations we suspect that it would generalise to alternative theories of gravity. It should also be noted that our results, in conjunction with the previous results of Hawking \cite{hawking72b} and Sotiriou \cite{sotiriou12}, completely exclude the existence of any new stationary black holes in the scalar-tensor theory of gravity, even allowing for time dependence.

There are several questions one might ask concerning our results. Maybe the most obvious one is are the conditions in Eq. \eqref{3.6} essential, or can they be weakened somewhat? More importantly, are there in fact any examples of stationary black hole solutions with time-dependent scalar fields at all? It is not in itself difficult to find examples of \(P\) which satisfies both of the conditions in Eq. \eqref{3.6}. A simple example is given by the choice \(P=(X-c)^{n}\), where \(n>1\) is an integer and \(c\) a positive constant; this satisfies both of the conditions in Eq. \eqref{3.6} with \(\alpha^{2}=2c\). Needless to say, it does not suffice to find a choice of \(P\) for which both conditions in Eq. \eqref{3.6} are satisfied for non-zero \(\alpha\), since we would need to demonstrate all other components of the Einstein equations could be satisfied and that it was indeed a black hole.

Another question one might consider is would these conclusions still hold if another matter field was added, for instance an electromagnetic field? While it is easy to see that the argument involving the Einstein equations in Eq. \eqref{3.4} would still hold provided the new matter field obeyed \(T_{tr}=0\) and \(T_{t\theta}=0\) we cannot necessarily conclude that \(\Phi\) is linear in time. This is because it is not required that both of the energy-momentum tensors separately have no time dependence; the Einstein equations only require that the sum of them has no time dependence. This is one of the reasons the argument only works for a single scalar field. 

We should also note that the argument crucially depended upon the spacetime being asymptotically flat (or some similar condition). In fact, exact static, spherically symmetric solutions of the Einstein equations with a massless, time-dependent scalar field are known and have been given by Wyman \cite{wyman81}. It would be interesting to find the analogue of Wyman's solutions when the scalar field is non-canonical. In this case one would only have to study the modified Einstein equations, since a scalar field profile of the form Eq. \eqref{3.45} always solves the scalar field equation of motion if the geometry is stationary and \(P_{,\Phi}=0\). 

One might also wonder about scalar fields which do not satisfy the null energy condition. A canonical scalar field must necessarily satisfy this, but it will be only satisfied for a non-canonical scalar field in general if \(P_{,X}>0\). In this case the rigidity theorem cannot be assumed, and so the metric in Eq. \eqref{2.1} may not be the most general stationary metric. We can, however, prove a version of this theorem which rules out static solutions with time-dependent scalar fields, without assuming any energy conditions. This is because in this case coordinates \((t,x^{i})\) may be chosen so that the spacetime metric takes the form
\begin{equation} \label{6.1} 
ds^{2}=g_{00}(\vec{x})dt^{2}+g_{ij}(\vec{x})dx^{i}dx^{j}, 
\end{equation}   
and an elementary calculation shows that \(R_{tx_{i}}=0\) for this metric (recall by the Cotton-Darboux theorem \cite{chandrasekhar} that the spatial coordinates may be chosen to diagonalise the metric). The off-diagonal Einstein equations will then imply that \(\phi=\phi(t)\), and the argument proceeds as before.

Finally, one might ask if this argument would work for Galileons and similar scalar field theories which explicitly depend upon more than one derivative of the scalar field \cite{nicolis09, horndeski74}. In such case the energy-momentum tensor will not be of the form of Eq. \eqref{3.3}, and in particular will likely explicitly depend upon two or more derivatives of the scalar field. We would therefore not expect the argument involving the Einstein equations to carry through automatically. In fact, some authors have recently found black hole solutions in these theories with time-dependent scalar hair \cite{babichev13, kobayashi14}. It is possible, though, that there exists some special choices of the action for which it works.

\section*{Acknowledgements}
A.A.H.G. is supported by the STFC. R.J. is supported by the Cambridge Commonwealth Trust and Trinity College, Cambridge. We thank John Barrow, Anne-Christine Davis, Sam Dolan and Elizabeth Winstanley for helpful discussions and comments on the draft.


\begin{thebibliography}{99}
 
\bibitem{wald84} 
R. M. Wald, \emph{General Relativity} (Chicago: U Chicago Press, 1984)

\bibitem{hawking73}
S. W. Hawking and G. F. R. Ellis, \emph{The Large Scale Structure of Space-Time} (Cambridge: Cambridge UP, 1973)

\bibitem{chandrasekhar}
S. Chandrasekhar, \emph{The Mathematical Theory of Black Holes} (Oxford: Oxford UP, 1983)

\bibitem{chrusciel12}
P. T. Chru\'sciel, J. L. Costa and M. Heusler, Stationary black holes: uniqueness and beyond, \emph{Living Rev. Relativity} {\bf 15}, 7 (2012) 

\bibitem{bekenstein98}
J. D. Bekenstein, Black holes: classical properties, thermodynamics, and heuristic quantization, arXiv: gr-qc/9808028

\bibitem{chase70}
J. E. Chase, Event horizons in static scalar-vacuum space-times, \emph{Commun. Math. Phys.} {\bf 19}, 276-288 (1970)

\bibitem{bekenstein72a}
J. D. Bekenstein, Transcendence of the law of baryon-number conservation in black hole physics, \emph{Phys. Rev. Lett.} {\bf 28}, 452-455 (1972)

\bibitem{bekenstein72b}
J. D. Bekenstein, Nonexistence of baryon number for static black holes, \emph{Phys. Rev. D} {\bf 5}, 1239-1246 (1972)

\bibitem{bekenstein72c}
J. D. Bekenstein, Nonexistence of baryon number for black holes. II, \emph{Phys. Rev. D} {\bf 5}, 2403-2412 (1972)

\bibitem{sudarsky95}
D. Sudarsky, A simple proof of a no-hair theorem in Einstein-Higgs theory, \emph{Class. Quantum Grav.} {\bf 12}, 579-584 (1995)

\bibitem{bekenstein95}
J. D. Bekenstein, Novel 'no-scalar-hair' theorem for black holes, \emph{Phys. Rev. D} {\bf 51}, R6608-R6611 (1995)

\bibitem{heusler96}
M. Heusler, \emph{Black Hole Uniqueness Theorems} (Cambridge: Cambridge UP, 1996)

\bibitem{graham14}
A. A. H. Graham and R. Jha, Non-Existence of black holes with non-canonical scalar fields, \emph{Phys. Rev. D} {\bf 89}, 084056 (2014)

\bibitem{hui13}
L. Hui and A. Nicolis, A no-hair theorem for the Galileon, \emph{Phys. Rev. Lett.} {\bf 110}, 241104 (2013)

\bibitem{sotiriou13}
T. P. Sotiriou and S-Y. Zhou, Black hole hair in generalized scalar-tensor gravity, \emph{Phys. Rev. Lett.} {\bf 112}, 251102 (2014)

\bibitem{hoenselaers78}
C. Hoenselaers, On the effect of motions on energy momentum tensors, \emph{Prog. Theor. Phys.} {\bf 59}, 1518-1521 (1978)

\bibitem{liebling12}
S. L. Liebling and C. Palenzuela, Dynamical boson stars, \emph{Living Rev. Relativity} {\bf 15}, 6 (2012)

\bibitem{herdeiro14}
C. A. R. Herdeiro and E. Radu, Kerr black holes with scalar hair, \emph{Phys. Rev. Lett.} {\bf 112}, 221101 (2014)

\bibitem{hawking72}
S. W. Hawking, Black holes in general relativity, \emph{Commun. Math. Phys.} {\bf 25}, 152-166 (1972)

\bibitem{hollands07}
S. Hollands, A. Ishibashi and R. M. Wald, A higher dimensional stationary rotating black hole must be axisymmetric, \emph{Commun. Math. Phys.} {\bf 271}, 699-722 (2007)

\bibitem{mukhanov99}
C. Armend\'ariz-Pic\'on, T. Damour and V. Mukhanov, \(k\)-Inflation, \emph{Phys. Lett. B} {\bf 458}, 209-218 (1999)

\bibitem{chiba00}
T. Chiba, T. Okabe and M. Yamaguchi, Kinetically driven quintessence, \emph{Phys. Rev. D} {\bf 62}, 023511 (2000)

\bibitem{ap00}
C. Armendariz-Picon, V. Mukhanov and P. J. Steinhardt, Dynamical solution to the problem of a small cosmological constant and late-time cosmic acceleration, \emph{Phys. Rev. Lett.} {\bf 85}, 4438-4441 (2000)

\bibitem{ap01}
C. Armendariz-Picon, V. Mukhanov and P. J. Steinhardt, Essentials of k-essence, \emph{Phys. Rev. D} {\bf 63}, 103510 (2001)

\bibitem{lim05}
C. Armendariz-Picon and E. A. Lim, Halos of K-essence, \emph{JCAP} {\bf 0508}, 007 (2005)

\bibitem{rendall06}
A. D. Rendall, Dynamics of k-essence, \emph{Class. Quantum Grav.} {\bf 23}, 1557-1570 (2006) 

\bibitem{wyman81}
M. Wyman, Static spherically symmetric scalar fields in general relativity, \emph{Phys. Rev. D} {\bf 24}, 839-841 (1981)

\bibitem{fisher48}
I. Z. Fisher, Scalar mesostatic field with regard for gravitational effects, \emph{Zh. Eksp. Teor. Fiz.} {\bf 18}, 636-640 (1948); arXiv:gr-qc/9911008

\bibitem{janis68}
A. I. Janis, E. T. Newman and J. Winicour, Reality of the Schwarzschild singularity, \emph{Phys. Rev. Lett.} {\bf 20}, 878-880 (1968)

\bibitem{torii99}
T. Torii, K. Maeda and M. Narita, Toward the no-scalar-hair conjecture in asymptotically de Sitter spacetime, \emph{Phys. Rev. D} {\bf 59}, 064027 (1999)

\bibitem{torii01}
T. Torii, K. Maeda and M. Narita, Scalar hair on the black hole in asymptotically anti-de Sitter spacetime, \emph{Phys. Rev. D} {\bf 64}, 044007 (2001)

\bibitem{hawking72b}
S. W. Hawking, Black holes in the Brans-Dicke theory of gravitation, \emph{Commun. Math. Phys.} {\bf 25}, 167-171 (1972)

\bibitem{sotiriou12}
T. P. Sotiriou and V. Faroni, Black holes in scalar-tensor gravity, \emph{Phys. Rev. Lett.} {\bf 108}, 081103 (2012)

\bibitem{nicolis09}
A. Nicolis, R. Rattazzi and E. Trincherini, Galileon as a local modification of gravity, \emph{Phys. Rev. D} {\bf 79}, 064036 (2009)

\bibitem{horndeski74}
G. W. Horndeski, Second-order scalar-tensor field equations in a four-dimensional space, \emph{Int. J. Theor. Phys.} {\bf 10}, 363-384 (1974)

\bibitem{babichev13}
E. Babichev and C. Charmousis, Dressing a black hole with a time-dependent Galileon, arXiv: 1312.3204 (2013)

\bibitem{kobayashi14}
T. Kobayashi and N. Tanahashi, Exact black hole solutions in shift symmetric scalar-tensor theories, \emph{Prog. Theor. Exp. Phys.} {\bf 2014}, 073E02 (2014)


\end{thebibliography}

\end{document}